\documentclass[12pt]{iopart}

\usepackage{graphicx}
\usepackage{amsfonts}
\usepackage{amssymb}
\usepackage{amsthm}
\usepackage{iopams}
\usepackage{color}
\usepackage{setstack}
\usepackage{dsfont}
 
\usepackage[colorinlistoftodos]{todonotes}

\newcommand{\ket}[1]{|{#1}\rangle}
\newcommand{\bra}[1]{\langle{#1}|} 

\begin{document}
\title{The roles of drift and control field constraints upon quantum control speed limits}
\author{Christian Arenz} 
\address{Frick Laboratory, Princeton University, Princeton NJ 08544, US}
\author{Benjamin Russell} 
\address{Frick Laboratory, Princeton University, Princeton NJ 08544, US}
\author{Daniel Burgarth} 
\address{Institute of Mathematics, Physics, and Computer Science, Aberystwyth University, Aberystwyth SY23 2BZ, UK}
\author{Herschel Rabitz}
\address{Frick Laboratory, Princeton University, Princeton NJ 08544, US}
\date{\today}

\begin{abstract}
In this work we derive a lower bound for the minimum time required to implement a target unitary transformation through a classical time-dependent field in a closed quantum system.
The bound depends on the target gate, the strength of the internal Hamiltonian and the highest permitted control field amplitude. 
These findings reveal some properties of the reachable set of operations, explicitly analyzed for a single qubit. 
Moreover, for fully controllable systems, we identify a lower bound for the time at which all unitary gates become reachable.
We use numerical gate optimization in order to study the tightness of the obtained bounds. 
It is shown that in the single qubit case our analytical findings describe  the relationship between the highest control field amplitude and the minimum evolution time remarkably well. 
Finally, we discuss both challenges and ways forward for obtaining tighter bounds for higher dimensional systems, offering a discussion about the mathematical form and the physical meaning of the bound.  
\end{abstract}

\maketitle

\section{Introduction}
Future and present quantum technologies, as well as experiments in highly sensitive quantum systems, require a fine degree of control over the considered system.
In particular, the preparation of states and the implementation of quantum gates for quantum information processing tasks both critically rely on high fidelity quantum operations.
It is vital to be able to implement such operations as accurately as possible with the available control resources, while also operating on a time scale significantly below the typical decoherence time scale of the system employed.
Quantum control, which is primarily focused on the task of `steering' a quantum system towards a desired target by using suitably tailored classical fields \cite{ControlRev1, ControlRev2}, has successfully been applied to a broad class of quantum systems for disparate purposes.
Diverse applications include: driving chemical reactions \cite{HerschMol}, entangling spin qubits in nitrogen vacancy centers \cite{SpinQubits} and noise filtering \cite{NoiseF}.
In general, much attention has been drawn to two aspects of quantum control theory, (i) the identification of the operations/states that can be implemented/prepared, and (ii) the calculation of corresponding pulses.
Regarding (i), the Lie theoretic approach, sometimes referred to as geometric quantum control theory, expresses the questions of reachability within the framework of Lie groups and Lie algebras \cite{BookDalessandro, BookJurdjevic, LieGeoTH}.
The frequently employed Lie algebra rank criterion \cite{LieRankC1} is a powerful tool which facilitates the determination of the reachable operations or states for a given quantum system steered by classical control fields.  
When it comes to the determination of the control fields (ii), both numerical and  analytical tools are used.
The deployment of optimal control theory \cite{BookDalessandro,OptimalC}, which is based on the Pontryagin maximum principle, can efficiently maximize the fidelity for reaching a desired target.
Typically this is done by numerically optimizing a given cost functional, sometimes subject to additional constraints, with a gradient based search \cite{Grape, DYNAMOpaper, GateFid2, exactgradient}.

While both the aforementioned aspects of quantum control have been extensively studied, much less attention has been devoted to understanding either the relevant time scales or the properties of the control fields necessary to implement a given target.
In various studies \cite{SpeedLimit1, SpeedLimit2, SpeedLimit3, SpeedLimit4, SpeedLimit5, SpeedLimit6, SpeedLimit8} lower bounds (known as \emph{quantum speed limits}, or \emph{QSL}), which characterize how fast a quantum system can evolve from an initial state to some final target state have been established for closed, finite dimensional systems. Additionally the QSL has been studied for implementing a two qubit gate using ultracold atoms in an optical potential \cite{SpeedLimit7}.  
In later works, these bounds have been extended to open systems \cite{SpeedLimitOpenS1, SpeedLimitOpenS2, SpeedLimitOpenS3, SpeedLimitOpenS5, SpeedLimitOpenS4}. Moreover, based on Lieb-Robinson bounds, speed limits for quantum information tasks such as the creation of entanglement were recently established \cite{SpeedLimitQuantumInfor}. 
In the closed system case, it has been shown that, for specific examples such limits can be reached by searching for control pulses using optimal control theory  \cite{OptimalControlSpeedLimit1, OptimalControlSpeedLimit2}. We remark here that this is a rather special case, while typically the standard QSL bounds are not tight when used for time dependent control systems, which will be discussed in section \ref{sec:tightness}. 

In cases in which the control fields are unconstrained, the minimum time (minimized over all pulses which implement a desired gate) to implement a target unitary transformation can be calculated analytically for simple models \cite{OptimalControlSpeedLimit2, ExactCalc1, ExactCalc2, ExactCalc3}.
Moreover, extensive numerical studies have been carried out to find the minimum gate time for more complex systems \cite{Carenzspinstar, Nori}.
However, finding the shortest possible control pulses in general remains challenging. It is broadly equivalent to finding geodesics of Randers type Finsler metrics on either (special) unitary groups  or  complex projective spaces for the tasks of implementing gates or preparing states, respectively \cite{Geo1, Geo2, Geo3, Geo4, Geo5}.
Other approaches also exist including brachistochrone equations \cite{Geo6}, however as of yet these can only be addressed by numerical approaches and they are not geometrically intrinsic rendering analytical solutions harder to obtain.
 
In light of the difficulty of finding general solutions to time optimal quantum control problems, it is desirable to establish lower bounds drawing on as much information as possible in order to obtain first estimates.
In particular, the effects of a constrained control field, the strength of the internal Hamiltonian and the choice of the target gate upon the minimum time are physically important and relatively unexplored.
Although detailed studies for qubit systems exist \cite{ReachQubit1, ReachQubit2, ReachMultiQubit, Ellie}, the characterization of the reachable set of operations as a function of the evolution time and constraints on either or both of the control field and the `internal' Hamiltonian also warrants further investigation.

The purpose of this article is twofold. The first objective is to address the aforementioned questions, while in a second step we will discuss the obstacles to obtaining more accurate bounds on minimum gate times. We emphasize that the standard QSL's, which are typically formulated for time independent systems, only depend on the geometry of the systems Hilbert space, whereas here we seek a QSL that is system and control dependent.  
The results serve as first estimates towards controlling complex quantum systems with feasible and robust pulses on an appropriate time scale.
Throughout this work we consider systems described by a Hamiltonian of the form
\begin{eqnarray}
\label{eq:ControleSG}
H(t)=H_{0}+f(t)H_{c},
\end{eqnarray}
where $f(t)$ is the control field and $H_{0}$ and $H_{c}$ are the drift and control Hamiltonians respectively.
Based on simple arguments, particularly an inequality from \cite{Geo1}, we derive for \eref{eq:ControleSG} a lower bound,
\begin{eqnarray}
  \label{eq:lowerboundtime}
  \frac{C(U_{g},H_{c})}{\Vert H_{0}\Vert}+\frac{C(U_{g},H_{0})}{|f_{\text{max}}|\Vert H_{c}\Vert  }\leq T,
\end{eqnarray}
for the time $T$ required to implement a target unitary operation $U_{g}$.
The quantities $C(U_{g},H_{c})$ and $C(U_{g},H_{0})$ depend on the target gate, the eigenbasis of the control and the drift Hamiltonian respectively, and the dimension of the quantum system being considered (see Eq. \eref{eq:defC1} and \eref{eq:defC2});  $f_{\text{max}}$ is the maximum permitted control field amplitude.
Since the bound depends on the target gate, for a given system we can further characterize the set of unitary operations which are not reachable for a fixed $f_{\text{max}}, ~T$ and $\Vert H_{0}\Vert, ~\Vert H_{c}\Vert$. We subsequently establish a lower bound for the time $T_c$ in which all gates become reachable as a corollary of this observation.
Numerical gate optimization using gradient ascent pulse engineering (GRAPE) is used in order to study the tightness of the obtained bounds. For a single qubit, we show that optimal control theory allows us to operate at the boundary (similar to a Pareto front) of the viable region in the $T,~f_{\text{max}}$ plane defined by \eref{eq:lowerboundtime}.
Finally, we discuss challenges to obtaining bounds which are tighter for higher dimensional systems, and further offer a discussion about the nature of the obtained bound \eref{eq:lowerboundtime}. 

\section{Bound on the minimum gate time}

We start by considering the following control system
\begin{eqnarray}
\label{eq:controlsystem}
\dot{U}(t)=-iH(t)U(t),~~~~U(0)=\mathds{1},	
\end{eqnarray}
on the unitary group $\text{U}(d)$ consisting of unitary $d\times d$ matrices.
Throughout this article we set $\hbar=1$.
We study Hamiltonians $H(t)$ of form \eref{eq:ControleSG}, in which the control field enters in a bilinear way, typically known as the \emph{dipole approximation} in chemical physics and as affine bi-linear control on Lie groups in the mathematics community \cite{Elliot}.
We further denote the set of gates, which can be reached at any time by some specific control field by $\mathcal R$.
It is well known that the closure of the reachable set $\overline{\mathcal R}$ is equal to the Lie group $e^{\mathfrak{L}}$ with $\mathfrak{L}=\text{Lie}(iH_{0},iH_{c})$ being the dynamical Lie algebra that is generated by iterated commutators and (real) linear combinations of the drift and the control Hamiltonian \cite{BookDalessandro}.
The system is said to be fully controllable iff $\mathfrak{L}=\mathfrak{u}(d)$ (or $\mathfrak{su}(d)$ for traceless Hamiltonians), where $\mathfrak{u}(d)$ is the Lie algebra of skew-hermitian matrices.  
Equivalently, for a fully controllable system every unitary goal gate $U_{g}\in \text{U}(d)$ can be implemented arbitrarily well \cite{BookDalessandro}. Remarkably, this is true for \emph{almost all} (all but a set of measure zero) control system of the form \eref{eq:ControleSG} \cite{BookJurdjevic, Altafini}.
For a more detailed introduction into quantum control theory and its terminology we refer to \cite{BookDalessandro,  BookJurdjevic, LieGeoTH}.
The dynamical Lie algebra is a powerful tool allowing one to identify the operations that can be implemented within a given control system.
However, it does not reveal anything about how much time is needed in order to implement a specific target, neither does it say anything about the strength of the corresponding control field(s).
Intuitively one would expect that if the strength of the control or the drift Hamiltonian decreases, then the time or the control field amplitude must correspondingly increase depending on the gate we want to implement.
In what follows we verify this intuition by establishing the lower bound \eref{eq:lowerboundtime}.

Using the triangle inequality, one can show that for two unitary operators $U_{1}(T)$ and $U_{2}(T)$, which are solutions to the Sch\"odinger equation at time $T$, the inequality \cite{Geo1} (see appendix  \ref{app:derivationNielsen} for a derivation), 
\begin{eqnarray}
\label{eq:NielsenBound}
\Vert U_{1}(T)-U_{2}(T)\Vert \leq \int_{0}^{T}\Vert H_{1}(t)-H_{2}(t)\Vert\,	dt
\end{eqnarray}
holds for any unitarily invariant norm with $H_{1}(t)$ and $H_{2}(t)$ being the Hamiltonians corresponding to the two trajectories $U_{1}(t)$ and $U_{2}(t)$ respectively.
Now, let $H_{1}(t)=H_{0}+f(t)H_{c}$ and $H_{2}(t)=f(t)H_{c}$ such that $U_{2}(T)=\exp(-i\alpha(T)H_{c})$ and $U_{1}(T)=U_{g}$ is the solution to \eref{eq:ControleSG} which implements the desired target and $\alpha(T)=\int_{0}^{T}f(t)dt$ is the integrated control field.
We  assume here that the target gate can be implemented by the given control system, i.e., $U_{g}\in \overline{\mathcal R}$, and we further note that any corresponding control field is not necessarily unique.
There can exist multiple different pulse shapes driving the system to the same target evolution for a given final time $T$.
Roughly speaking, Eq. \eref{eq:NielsenBound}, instantiated with the above choice for $U_{1}$ and $U_{2}$, yields a description of how much the drift Hamiltonian is ``needed'' in order to reach a gate. We note that a similar separation has been suggested in \cite{SpeedLimit8} by constructing an observable that commutes with $H_{0}$. However, the speed limit in \cite{SpeedLimit8} applies only to state-to-state transfer and it captures only the effect of constrained control fields and fails to characterize the role of the strength of the drift Hamiltonian. In the following we derive a speed limit for implementing a unitary transformation that explicitly incorporates the strength of the drift Hamiltonian, as well the maximum control field amplitude. We begin with the above choice for $U_{1}$ and $U_{2}$ in order to obtain a speed limit that depends on the strength of the drift Hamiltonian. Afterwards, we chose $U_{1}$ and $U_{2}$ differently in order to obtain another speed limit that depends on the maximum strength of the control field. A linear combination of both bounds yields the desired result \eref{eq:lowerboundtime} from the introduction. Evaluating \eref{eq:NielsenBound} for the Frobenius norm $\Vert A\Vert=\sqrt{\text{tr}\{A^{\dagger}A\}}$, which is used throughout this work, we find 
\begin{eqnarray}
\label{fundermentalultimatebound}
	\frac{\sqrt{2(d-\Re[\text{tr}\{U_{2}^{\dagger}(T)U_{g}\}])}}{\Vert H_{0}\Vert }\leq T,
\end{eqnarray}
and since $\Re[\text{tr}\{U_{2}^{\dagger}(T)U_{G}\}]\leq \sum_{j}^{d}|\bra{\phi_{j}^{(c)}}U_{g}\ket{\phi_{j}^{(c)}}|$ with $\{\ket{\phi_{j}^{(c)}}\}_{j=1}^{d}$ being the eigenbasis of $H_{c}$, we further conclude
\begin{eqnarray}
\label{eq:speedlimit}
\frac{\sqrt{2(d-\sum_{j}^{d}|\bra{\phi_{j}^{(c)}}U_{g}\ket{\phi_{j}^{(c)}}|)}}{\Vert H_{0}\Vert} \leq T.
\end{eqnarray}
Similar to the lower bounds that were obtained in \cite{SpeedLimit1, SpeedLimit2, SpeedLimit3}, the above inequality is a lower bound for the least time needed to implement a given target unitary gate. Henceforth, we refer to this time as the minimum gate time. The speed with which a desired given unitary can be implemented is inherently limited by the speed with which the propagator $U(t)$ evolves under the free evolution alone.
Unless one wants to implement a gate that can be reached by the control and $H_{c}$ alone, which can be done instantaneously if we assume that the amplitude of the control field is unconstrained, the strength of the drift Hamiltonian sets an ``intrinsic'' limit on how fast we can reach the desired target.
However, typically the amplitude of the control field is limited in any experimental situation, and therefore a practical lower bound for $T$ must also depend on the highest control field amplitude $f_{\text{max}}$.
Analogous to the derivation of \eref{eq:speedlimit}, but now with $H_{2}=H_{0}$, this can be established by using $\int_{0}^{T}\Vert f(t)H_{c}\Vert\,dt\leq T |f_{\text{max}}|\Vert H_{c}\Vert$.
We find
\begin{eqnarray}
\label{eq:speedlimitfield}
	\frac{\sqrt{2(d-\sum_{j}^{d}|\bra{\phi_{j}^{(0)}}U_{g}\ket{\phi_{j}^{(0)}}|)}}{|f_{\text{max}|}\Vert H_{c}\Vert} \leq T,
\end{eqnarray}
where $\{\ket{\phi_{j}^{(0)}}\}_{j=1}^{d}$ is the eigenbasis of $H_{0}$. Conversely to \eref{eq:speedlimit}, the bound \eref{eq:speedlimitfield} represents a speed limit that is enforced by limitations of the control field (extrinsic), rather than intrinsic limitations given by the strength of the drift Hamiltonian.
We postpone the discussion about the distinction between extrinsic and intrinsic speed limits to section \ref{ref:discussionextrinsicintrinsic} and proceed by defining
\begin{eqnarray}
\label{eq:defC1}
C(U_{g},H_{c})&\equiv \frac{\sqrt{2(d-\sum_{j}^{d}|\bra{\phi_{j}^{(c)}}U_{g}\ket{\phi_{j}^{(c)}}|)}}{2},\\
\label{eq:defC2}
 C(U_{g},H_{0})&\equiv	\frac{\sqrt{2(d-\sum_{j}^{d}|\bra{\phi_{j}^{(0)}}U_{g}\ket{\phi_{j}^{(0)}}|)}}{2}.
\end{eqnarray}
From \eref{eq:speedlimit} and \eref{eq:speedlimitfield} we then find $2C(U_{g},H_{c})/\Vert H_{0}\Vert+2C(U_{g},H_{0})/(|f_{\text{max}}|\Vert H_{c}\Vert)\leq 2T$. As such, the lower bound from the introduction \eref{eq:lowerboundtime} is obtained by linearly combining \eref{eq:speedlimit} and \eref{eq:speedlimitfield}. We note that each term of the left-hand side of \eref{eq:lowerboundtime} is weighted in a different manner by the target operation.
As described in \cite{meijfcs}, there are many ways to combine two or more speed limit formulas to create novel ones. As in \cite{MaxSpeedlim}, simply taking the  maximum of \eref{eq:speedlimit} and \eref{eq:speedlimitfield} yields $T~\geq~\max\{2C(U_{g},H_{c})/\Vert H_{0}\Vert,~2C(U_{g},H_{0})/(|f_{\text{max}}|\Vert H_{c}\Vert)\}$.  
Another method is to take convex combinations, which is done in this work using an equal weighting. The authors have not as of yet determined the combination that produces the tightest bound. 
However, the numerical simulations in section \ref{ref:pareto} suggest that this choice is worth investigation and moreover, it is conjectured in section \ref{sec:tightness} that speed limits for the control system \eref{eq:ControleSG} generally should be of this form. 
To summarize, the inequality \eref{eq:lowerboundtime} can be considered as a necessary condition which must be satisfied by $\Vert H_{0}\Vert,~\Vert H_{c}\Vert,~f_{\text{max}}$ and $T$ in order to implement some $U_{g}\in\overline{\mathcal R}$.

\subsection{Characterization of the reachable set}
Since the lower bound \eref{eq:lowerboundtime} depends on the target unitary transformation $U_{g}$, it reveals some information about the set of gates $\mathcal G_{T}$, which \emph{provably} cannot be reached for a given evolution time $T$.
For instance, consider the simplified case of implementing a gate that can be reached by the control Hamiltonian alone, i.e. $U_{g}=\exp(-i\alpha(T)H_{c})$. This yields $C(U_{g},H_{c})=0$.
Here the drift Hamiltonian is not required to reach the target evolution. However, if the control field is not sufficiently large the gate cannot be implemented. 
\subsubsection{Single qubit case:}
\begin{figure}[h!] \begin{minipage}[hbt]{7.0cm} a)\\
\includegraphics[width=1\columnwidth]{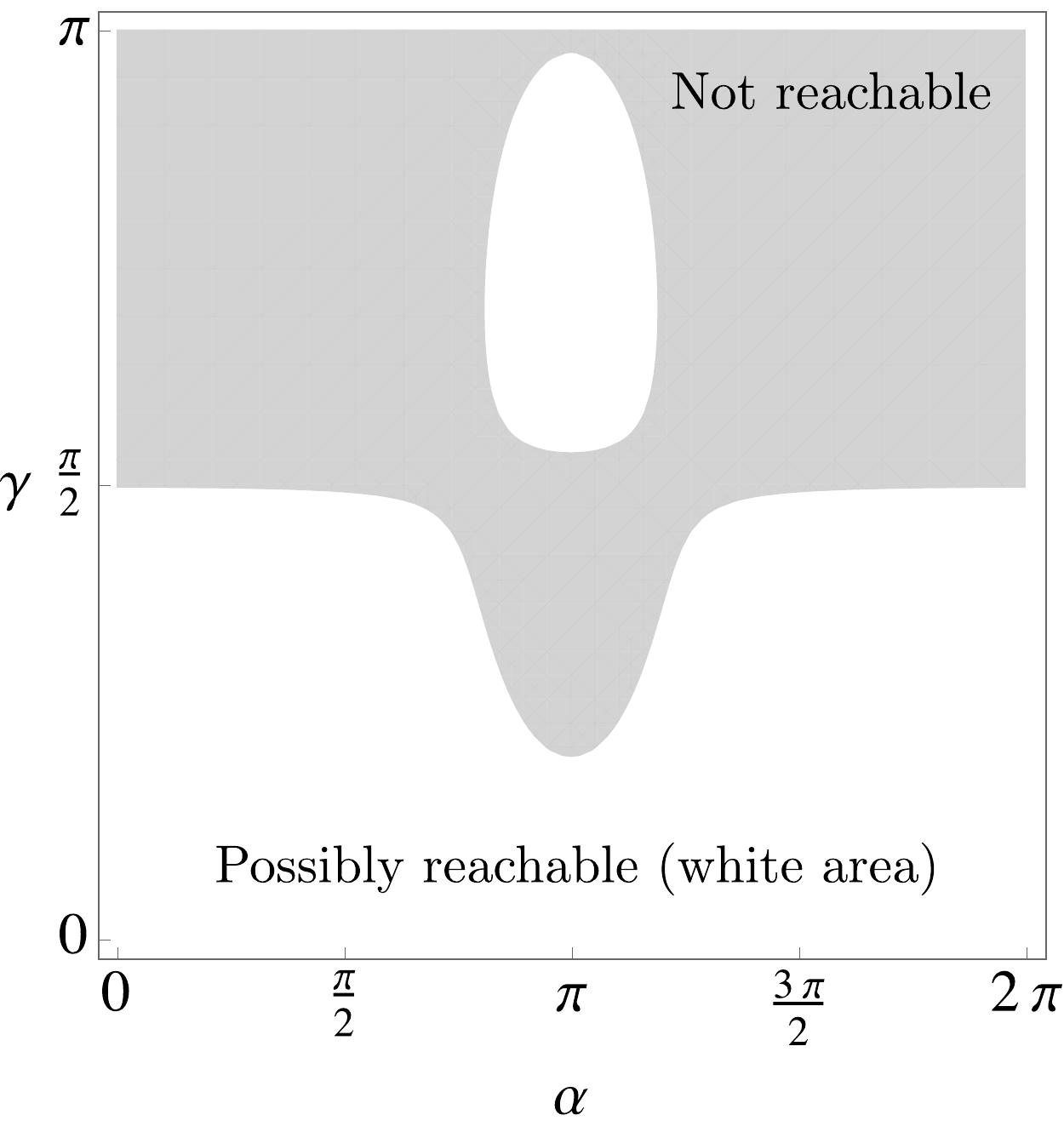} \end{minipage}
\hfill \begin{minipage}[hbt]{8.0cm} b)\\
\includegraphics[width=1\columnwidth]{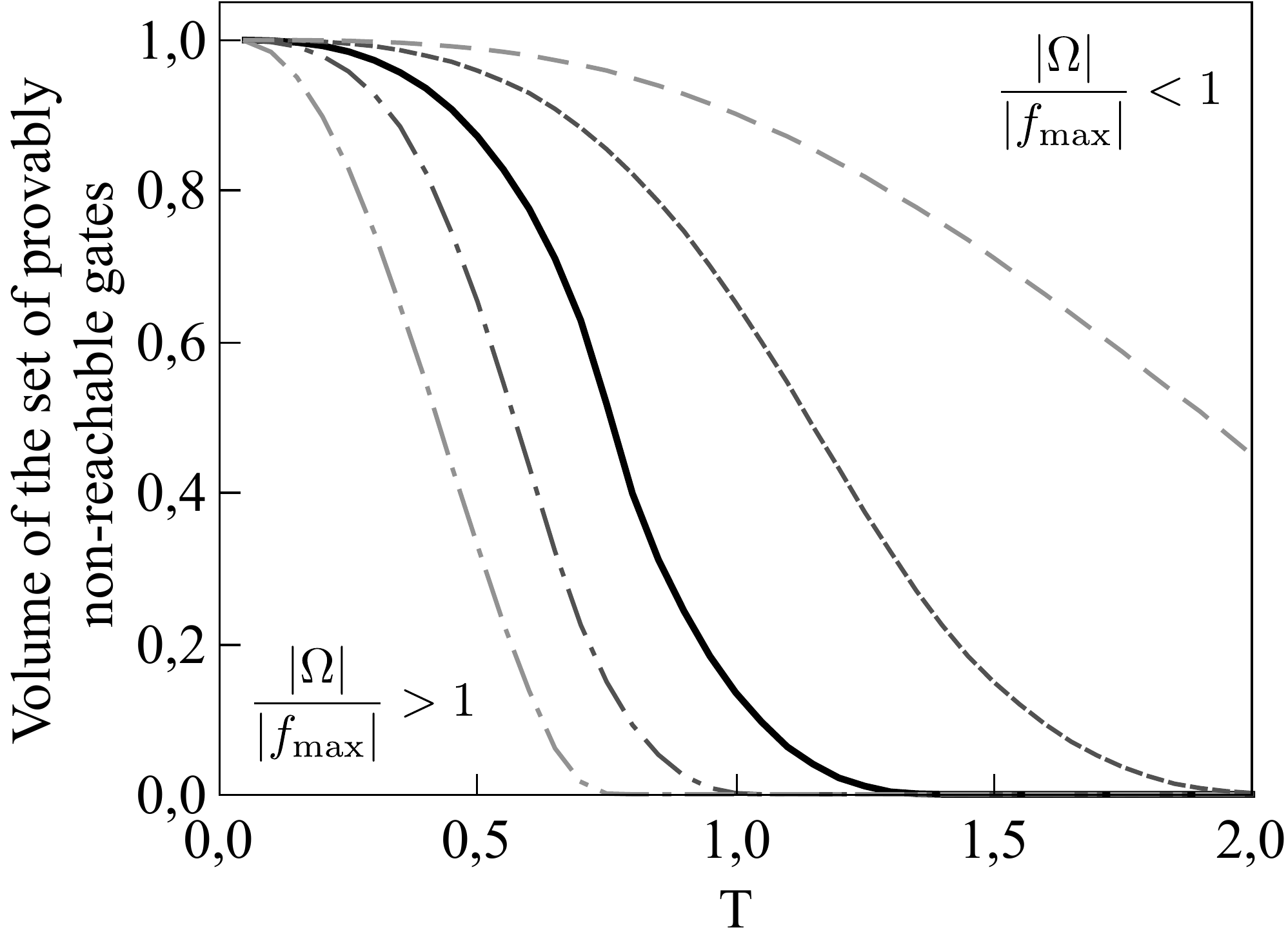}
\end{minipage} \caption{\label{fig:Qubit} Characterization of the gates that proveably cannot be reached for a single qubit described by the control system \eref{eq:singlequbit}. Based on \eref{eq:reachablesstatesqubit}, a) shows the set of states that cannot be reached (grey area) from an initial eigenstate of $\sigma_{z}$ for an evolution time $T=0.52$ and $\Omega=f_{\text{max}}=1$. Fig. b) shows the volume of gates (\eref{eq:volumedef} and \eref{eq:intregion}) that cannot be reached as a function of the evolution time $T$. The ratio $\frac{|\Omega|}{|f_{\text{max}}|}\in\{10,~2,~1,~0.5,~0.25\}$ was chosen from left to right in decreasing order, whereas the solid black line represents $\Omega=f_{\text{max}}=1$.}
\end{figure}

in order to study more complex cases requiring an interplay between the drift and the control Hamiltonian, we consider a single qubit described by the Hamiltonian
\begin{eqnarray}
\label{eq:singlequbit}
H(t)=\Omega \sigma_{x}+f(t)\sigma_{z},	
\end{eqnarray}
where $\sigma_{j}$, with $j=x,y,z$, are the Pauli matrices. We remark that the reachable set of a single qubit subject to two independent control fields was recently analyzed in great detail \cite{ReachQubit1, ReachQubit2}. The system is fully controllable hence every $U_{g}\in \text{SU}(2)$ can be implemented. We can parameterize a $U_{g}=R_{z}(\alpha)R_{y}(\gamma)R_{z}(\beta)$ with the three angles $0\leq \alpha < 2\pi,~0\leq \beta < 4\pi,~0\leq \gamma \leq \pi$ where $R_{z}(\alpha)=\exp(-i\frac{\alpha}{2} \sigma_{z})$ and $R_{y}(\gamma)=\exp(-i\frac{\gamma}{2} \sigma_{y})$ are rotations around $\sigma_{z}$ and $\sigma_{y}$ respectively such that $C(U_{g})=C(\alpha,\beta,\gamma)$. For an unconstrained control field, the time required to implement some $U_{g}\in \text{SU}(2)$ can be calculated exactly \cite{OptimalControlSpeedLimit2} using the Euler angle decomposition above. This calculation shows that the minimum time is determined by $\Omega$. With the established bound \eref{eq:lowerboundtime}, we can now proceed with analyzing the effect of a constrained control field amplitude. For $\beta=0$, i.e when every state can be reached from an initial eigenstate of $\sigma_{z}$, the lower bound takes the form  
\begin{eqnarray}
\label{eq:reachablesstatesqubit}
\frac{\sqrt{(2-2\cos(\gamma/2))}}{4|\Omega|}+\frac{\sqrt{3-\cos(\alpha)\cos(\gamma)}}{4|f_{\text{max}}|}\leq T.
\end{eqnarray}
Fig. \ref{fig:Qubit} a) shows the set of states, parametrized through $\gamma$ and $\alpha$, that do not satisfy \eref{eq:reachablesstatesqubit} (grey area) for a fixed evolution time $T=0.53$ and $\Omega=f_{\text{max}}=1$, hence these states cannot be reached. The white area contains all states that do satisfy \eref{eq:reachablesstatesqubit}, nonetheless, this does not reveal whether they are reachable or not. This question is related to the tightness of the bound \eref{eq:reachablesstatesqubit}, which will be analyzed in the next section \ref{ref:pareto}.  

A way to study which gates are provably not reachable as a function of $T$ for a given $\Omega,~f_{\text{max}}$ is to consider the volume of the set $\mathcal G_{T}$. For $\text{SU}(2)$, the measure such that the volume of the entire group is one, is given by $dV=\frac{1}{16\pi^{2}}\sin(\gamma)\,d\gamma\,d\alpha\,d\beta$. As such, for a single qubit the volume $V(\mathcal G_{T})$  can be calculated as
\begin{eqnarray}
\label{eq:volumedef}
V(\mathcal G_{T})=\frac{1}{16\pi^{2}}\int_{\Sigma(\mathcal G_{T})}\sin(\gamma)\,d\gamma\,d\alpha\,d\beta,	
\end{eqnarray}
where  
\begin{eqnarray}
\label{eq:intregion}
\Sigma(\mathcal G_{T})=\left\{(\alpha,\beta,\gamma)\,\Big|\, \frac{C(\alpha,\beta,\gamma,\sigma_{z})}{\sqrt{2}|\Omega|}+\frac{C(\alpha,\beta,\gamma,\sigma_{x})}{\sqrt{2}|f_{\text{max}}|}\leq T      \right\},	
\end{eqnarray}
is the integration region. In Fig. \ref{fig:Qubit} b) we numerically integrated \eref{eq:volumedef} for different values $T$. The solid black line shows the case where $\Omega=f_{\text{max}}=1$ and the dashed lines (dashed-dotted lines) show the cases where $|\Omega|<|f_{\text{max}}|$ ($|\Omega|>|f_{\text{max}}|$) with $f_{\text{max}}=1$. 

\subsubsection{The time required to implement all gates:} from Fig. \ref{fig:Qubit} we observe that when the evolution time becomes larger the number of gates that proveably \emph{cannot} be implemented becomes monotonically \emph{smaller}.
It is known (Theorems 1 and 3 \cite{BookJurdjevic}, which lead to the result in \cite{qclcl}) that for a fully controllable system there exist a time $T_{c}$ for which all gates can be implemented.
In order to establish a lower bound for $T_{c}$ we seek the gate for which $C(U_{g},H_{0})$ and $C(U_{g},H_{c})$ become maximal. 
In the single qubit case \eref{eq:singlequbit} this can be achieved simply through $U_{g}=\sigma_{y}$.
Unfortunately, unless $H_{0}$ and $H_{c}$ have eigenbasis which are mutually unbiased \cite{MUTU}, finding maximizing $U_{g}$ in general remains an open problem.
However, for a fully controllable qubit system there always exists a $U_{g}$ (see \ref{app:derivationLB} for further details) for which $\Re[\text{tr}\{U_{g}^{\dagger}U^{(i)}\}]\leq d/2$ with $i=1,2$ where $U^{(1)}=\exp(-i\alpha(T_{c})H_{c})$ and $U^{(2)}=\exp(-iT_{c}H_{0})$.
With $\Vert H_{0}\Vert\leq \sqrt{d}|E_{0}|$ and $\Vert H_{c}\Vert\leq \sqrt{d}|E_{c}|$,  where $E_{0}$ and $E_{c}$ are the highest eigenvalue of $H_{0}$ and $H_{c}$ respectively, we thus have 
\begin{eqnarray}
\label{eq:criticaltimebound}
\frac{1}{2|E_0|}+\frac{1}{2|f_{\text{max}}E_c|} \leq T_{c}.
\end{eqnarray}
For $E_0,~E_c,$ and $f_{\text{max}}$ determined by the quantum system and experimental control limitations, respectively, the bound \eref{eq:criticaltimebound} determines how at least much time is in order to be able to implement all gates.
In particular, an obstacle to fully controlling the system on a implementable time scale arises when the norm of $H_{0}$ decreases with an increase of the dimension of the system, or, when $f_{\text{max}}$ is not sufficiently large.

In order to be able to implement all gates, $E_0$, $E_{c}$ and $f_{\text{max}}$ must be given in such a way that $T_{c}$ does not reach an order within which other effects, such as decoherence, cannot be neglected. 
Denoting by $T_{\text{Dec}}$ the typical decoherence time scale, $|2E_0|^{-1}+|2f_{\text{max}}E_c|^{-1}\leq T_{\text{Dec}} $ needs to be satisfied in order to be able to implement all gates.
We note here that this is a heuristic argument rather than a rigorous conclusion as the application of control fields can substantially change the effect of the environment \cite{Kurizki}. For example, in the extreme case of an infinitely fast decoupling sequence the effect of the environment can be completely suppressed for a large class of system-environment interactions \cite{LVioal1, Dec1me}. In such a way coherence times can be significantly prolonged \cite{QubitinSolid}.
Opposingly, as shown in \cite{SchmidtCalarco, KochReich2, Me}, sometimes the environment and noise that is caused by it are beneficial, even turning the system into a fully controllable one \cite{Me}. A detailed and rigorous analysis, including the interaction with an environment, is beyond the scope of this work and will be the subject of future studies.

\subsection{Tightness of the bound and Pareto optimal control}
\label{ref:pareto}
As mentioned in the introduction, the aim of optimal control theory is to find a pulse that maximizes or minimizes a given cost functional. For the implementation of a target unitary gate $U_{g}$ this is typically done by minimizing the infidelity 
\begin{eqnarray}
\label{eq:infidelity}
\epsilon=1-\left|\frac{1}{d}\text{tr}\{U_{g}^{\dagger}U(T)\}\right|^{2},
\end{eqnarray}
using a gradient based search \cite{Grape, DYNAMOpaper, GateFid2, exactgradient}, such as the GRAPE algorithm \cite{Grape}.
In order to study the tightness of the bounds \eref{eq:lowerboundtime} and \eref{eq:speedlimit} we employ in this section numerical minimization of the infidelity $\epsilon$ using the GRAPE algorithm in the QuTip control package \cite{QuTip1, QuTip2}.
We begin by analyzing the bound \eref{eq:speedlimit}, i.e., the case in which the control field amplitude is not constrained such that the speed limits only arise from the limited strength of the drift Hamiltonian.
We study the single qubit example from the previous section for a target evolution $U_{g}=\sigma_{y}$ and an N-level system which is known to be fully controllable \cite{FullControlSingleAcc}.
The drift Hamiltonian reads $H_{0}=J\sum_{j=1}^{N-1}(\ket{j}\bra{j+1}+\text{h.c.})$ where control is exerted trough $H_{c}=\ket{1}\bra{1}$ and as a target evolution we consider the SWAP gate $U_{g}=\exp(-i\pi/2(\ket{1}\bra{N}+\ket{N}\bra{1}))$. 

\begin{figure} \begin{minipage}[hbt]{8.0cm} a)\\
\includegraphics[width=1.0\columnwidth]{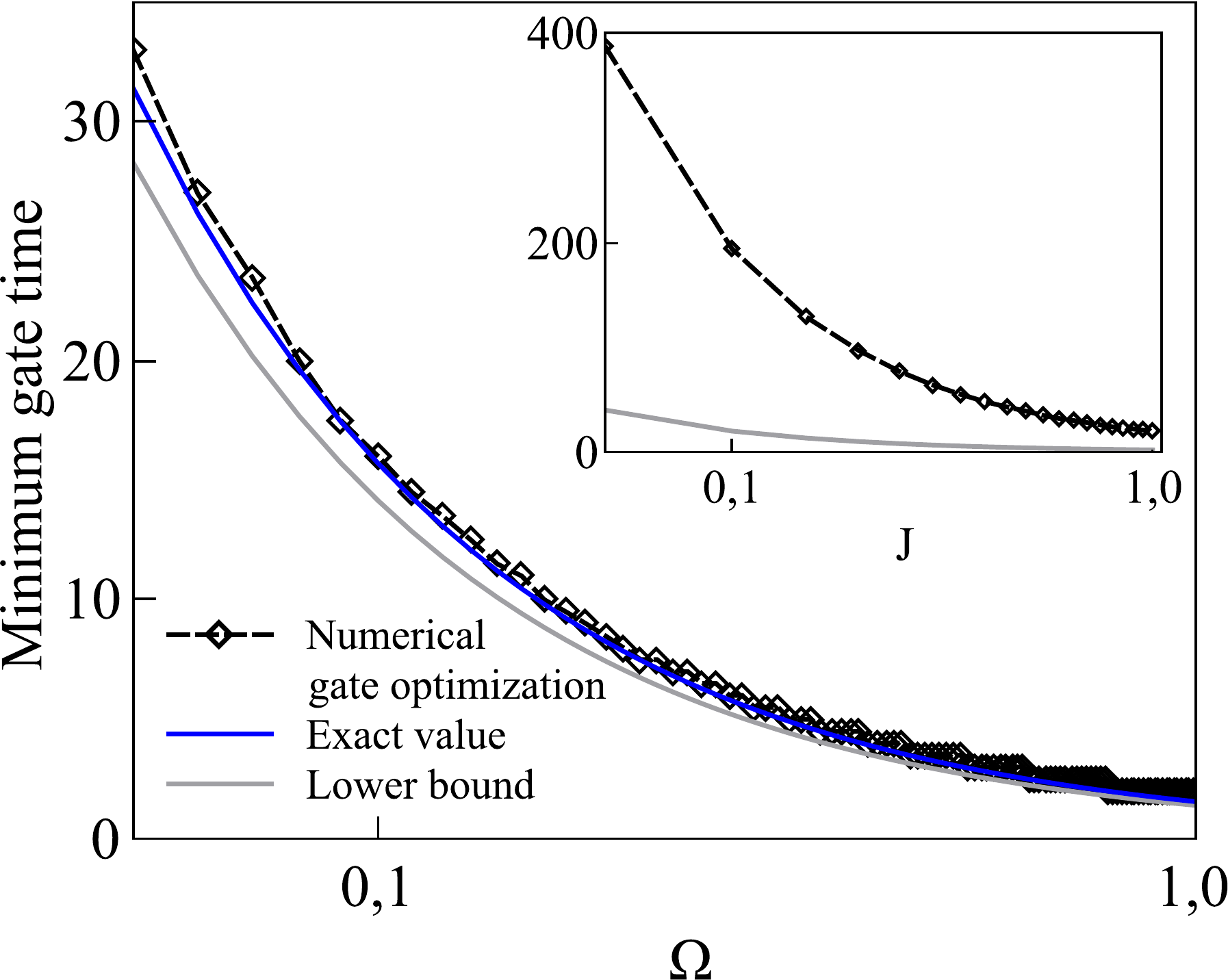} \end{minipage}
\hfill \begin{minipage}[hbt]{6.5cm} b)\\
\includegraphics[width=1.0\columnwidth]{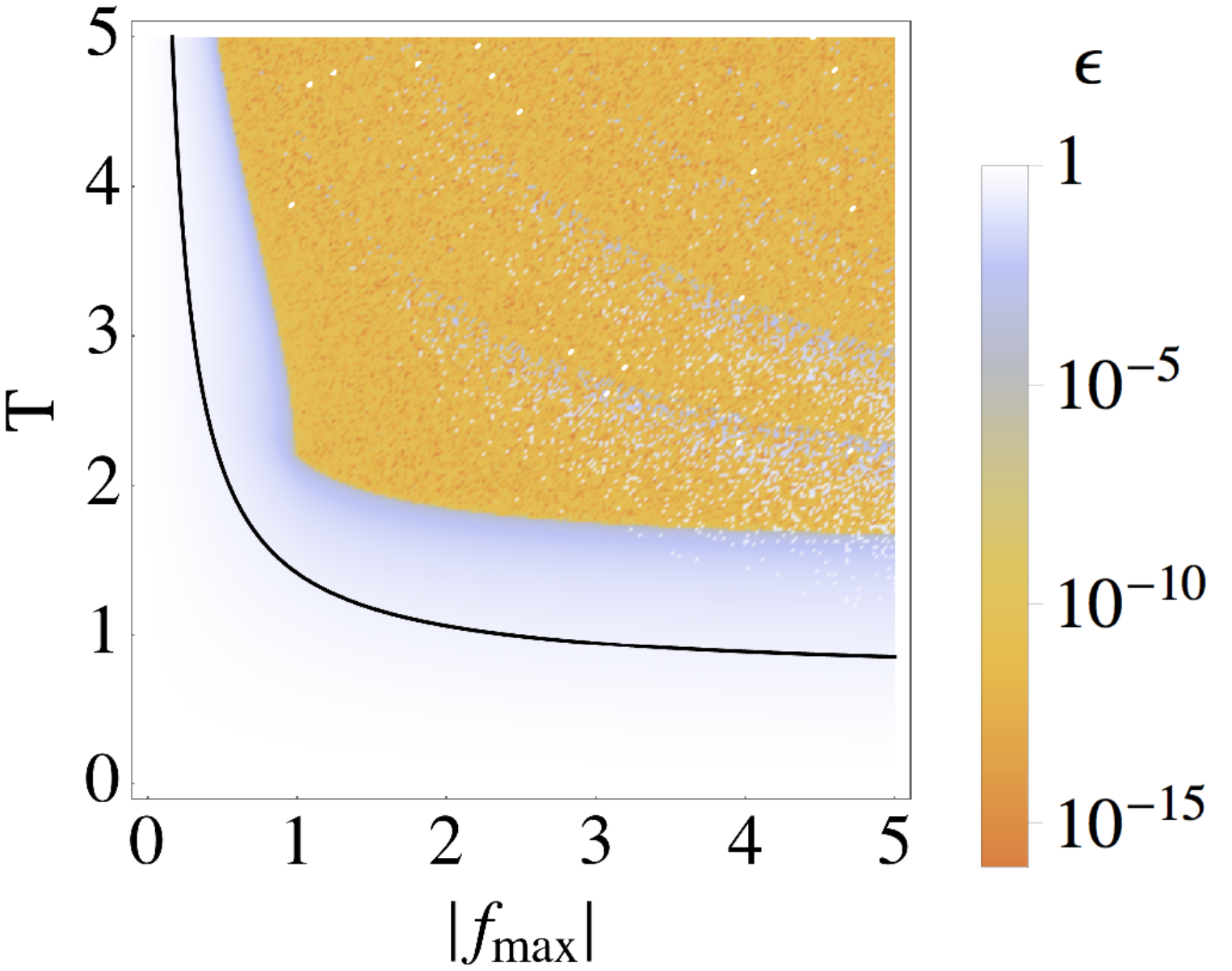}
\end{minipage} \caption{\label{fig:ParetoNum} Numerical gate optimization of the infidelity $\epsilon$ given by \eref{eq:infidelity} to study the tightness of the bounds \eref{eq:lowerboundtime} (grey curves) and \eref{eq:speedlimit} (black curve) shown in a) and b) respectively. a) Minimum gate time as a function of the strength $\Omega$ and $J$ of the drift Hamiltonian on a logarithmic scale for the single qubit control system \eref{eq:singlequbit} with a target evolution $U_{g}=\sigma_{y}$ and (inset plot) a N-level system (details can be found in the main text) with $N=4$ levels and a SWAP gate as a target evolution.
The diamonds show the numerically obtained values and in the single qubit case the exact value of the minimum gate time (blue curve) is given by $T^{*}=\frac{\pi}{2|\Omega|}$. b) Numerical gate optimization of the infidelity for different values of the evolution time $T$ and the highest permitted control field amplitude $f_{\text{max}}$ evaluated for a single qubit \eref{eq:singlequbit} with $U_{g}=\sigma_{y}$ and $\Omega=1$.
The colormap shows the infidelity, where in the orange region, $\epsilon<10^{-10}$ is achieved.}
\end{figure}

In Fig. \ref{fig:ParetoNum} a) we show the minimum gate time as a function of the norm of the drift Hamiltonian, where the inset plot shows the N-level system with $N=4$ levels and $H_{0}$ was normalized in such a way that $\Vert H_{0}\Vert=J$.
In both cases the grey curves represent the lower bound \eref{eq:speedlimit} and the numerically estimated values for the minimum gates times (black diamonds) were obtained by minimizing $\epsilon$ for different values of the total evolution time $T$ until a threshold of $\epsilon< 10^{-7}$ is reached.
As mentioned in the previous section, for the single qubit control system \eref{eq:singlequbit} with an unbounded control field amplitude the minimum gate time $T^{*}$ can be calculated exactly \cite{OptimalControlSpeedLimit2}, yielding for $U_{g}$ from above $T^{*}=\frac{\pi}{2|\Omega|}$ (blue curve in Fig. \ref{fig:ParetoNum} a)).
We emphasize that the numerically obtained values are themselves only an upper bound since the convergence of the optimization algorithm depends on the initial trial pulse, which was chosen randomly in all cases.
Nevertheless, from Fig. \ref{fig:ParetoNum} a) we conclude that, remarkably, the lower bound \eref{eq:speedlimit} is tight for the single qubit control system.
Unfortunately, as indicated by the inset in Fig. \ref{fig:ParetoNum} a), this is less satisfactory when the dimension of the quantum system increases.

Besides minimizing $\epsilon$, sometimes there are additional constraints one must take into account. For instance, one wants to find the optimal control pulses that minimize $\epsilon$, while keeping the length of the pulses as short as possible and additionally using the least amount of energy. Such a multi-objective optimization is known as Pareto optimal control \cite{BookHillermeier, BookStatMat, Hersch}. Typically one seeks to identify non-dominated solutions, and it is generally not possible to achieve fully optimal solutions that maximize all objectives \cite{Hersch}.   
We consider here the situation where we (i.e., $\epsilon=0$) want to perfectly implement some $U_{g}$ in the shortest possible time $T$ while simultaneously constraining the highest control field amplitude $f_{\text{max}}$ as much as possible. Clearly, there is a trade off between the three objectives, meaning that we lose fidelity when we constrain either the control field or the evolution time too much. A general question is how much we can reduce the length and the highest amplitude of the pulse, while still being able to implement $U_{g}$. Rearrangement of \eref{eq:lowerboundtime} yields
\begin{eqnarray}
\label{eq:boundTF}
T-\frac{1}{|f_{\text{max}}|}\frac{C(U_{g},H_{0})}{\Vert H_{c}\Vert}\geq \frac{C(U_{g},H_{c})}{\Vert H_{0}\Vert},	
\end{eqnarray}
which defines a region that characterizes how much $T$ and $f_{\text{max}}$ can be minimized while still begin able to implement some $U_{g}$. Clearly, the inequality \eref{eq:boundTF} is a lower bound for the minimization over $T$ and $f_{\text{max}}$, whereas the actual smallest values for which the infidelity $\epsilon$ is still zero might be larger. In the following we therefore want to analyze the tightness of the lower bound \eref{eq:boundTF}. Again, we resort to numerical gate optimization of the infidelity $\epsilon$, but now for different values of $T$ and $f_{\text{max}}$. We focus on a single qubit described though the control system \eref{eq:singlequbit} with $\Omega=1$. As a target evolution we again take $U_{g}=\sigma_{y}$, which yields $T-1/(\sqrt{2}|f_{\text{max}}|)\geq 1/(\sqrt{2})$. The results are shown in Fig. \ref{fig:ParetoNum} b) wherein the black curve represents the lower bound and the orange area represents the achievement of infidelities of $\epsilon<10^{-10}$. The points that seem to break the continuity in the orange region are numerical artifacts, which can be removed by additionally minimizing over different initial pulses. From Fig. \ref{fig:ParetoNum} b) we observe that the bound \eref{eq:boundTF} is tight too for the single qubit control system, but, as before, similar simulations for higher dimensional systems indicate that tightness is lost.

\section{Discussion}

The bound \eref{eq:lowerboundtime} has been shown to be an excellent approximation to the numerically obtained values in the case of a single qubit system. However, for higher dimensional systems, the bound becomes diminishingly tight. In the following we want to discuss the reasons that this appears to be the case and a way forward to obtain tighter bounds for higher dimensional systems. Moreover, we make the distinction between two types of bounds for the minimum gate time, to which we refer as \emph{intrinsic} and \emph{extrinsic}. They arise from limitations of the drift or the control Hamiltonian (intrinsic) or some limited control resources (extrinsic), such as constrained control fields.

\subsection{Tightness}
\label{sec:tightness}
The bound \eref{eq:lowerboundtime} can be considered as a \emph{first order} approximation to the true minimum gate times.
The approximating step can be traced back to the use of the triangle inequality in the derivation of \eref{eq:NielsenBound} (see \ref{app:derivationNielsen}, Eq. \eref{eq:usetriangle}).
In this step higher order commutator expressions contributing to the trajectory of the unitary propagator $U(t)$ are disregarded.
By observing the nature of the terms in \eref{eq:lowerboundtime}, we see that while $\Vert H_0 \Vert$ (or $\Vert H_{c}\Vert$) and the eigenbasis of $H_c$ ($H_{0}$) both appear, the norm of the commutator $[iH_0, iH_c]$ does not appear.
Furthermore, all additional commutator terms of $iH_0$ and $iH_c$ are also absent.
For controllable systems, the set of all such nested commutator expressions \cite{Altafini} must generate the whole algebra $\mathfrak{u}(d)$.
For a visual example of the way in which such bracket expressions appear, see the `Lie tree' diagrams in \cite{Carenzspinstar}.
As such, a critical part of the dynamics of a system evolution is disregarded by any bound on minimum gate times (or any other QSL formula used in quantum control) which does not take these additional commutator terms, that is the structure of the underlying dynamical Lie algebra, into account.
In the same work numerical gate optimization suggests that the minimum gate time for a specific model scales exponentially with the number of qubits. 
As the dimension of a system rises, the nested commutator depth required to span the full algebra $\mathfrak{u}(d)$ grows \cite{Carenzspinstar, Elliot}.
The authors conjecture that tighter bounds on minimum gate times in terms of maximum control field amplitudes can be obtained by incorporating higher order terms of nested commutator expressions. As such, the authors anticipate the possibility of establishing tighter bounds of the form 
\begin{eqnarray}
\label{eq:formtigherbound}
\frac{L^{\bot}}{\Vert H_{0}\Vert}+\frac{L}{|f_{\text{max}}|\Vert H_{c}\Vert}\leq T, 
\end{eqnarray}
where $L$ and $L^{\bot}$ are two contributions to the length of the time optimal trajectory connecting $U(0)=\mathds{1}$ and $U(T)=U_{g}$. In \cite{ExactCalc1, ExactCalc2} it was shown that the group is covered (formally, foliated) by a family of subsets, known as \emph{cosets}, which play a crucial role in characterizing the time optimal trajectories in systems with unbounded controls. These subsets are related to the unitary operations corresponding to the control Hamiltonians alone; evolutions of arbitrarily high speed are possible within these sets using the controls alone, provided that the controls are unconstrained.  Moreover, the total length of any time optimal trajectory splits into two contributions. Firstly, $L^{\bot}$, the length of the trajectory between the cosets (i.e., orthogonal to each coset) and secondly, $L$, the length of the trajectory within cosets. The speed at which a time optimal trajectory can be traversed also splits into two parts, namely $\Vert H_{0}\Vert$, the speed between cosets and,  the speed of the evolution within cosets. For a constrained control field this speed is bounded by $|f_{\text{max}}|\Vert H_{c}\Vert$. Hence tighter bounds than \eref{eq:lowerboundtime} are expected to be still of the form \eref{eq:formtigherbound}.

\subsection{Intrinsic and extrinsic speed limits}
\label{ref:discussionextrinsicintrinsic}
Many works have recently focused on determining minimum gate times, or quantum speed limits.
We emphasize here two clear types of bounds which are in regular use, but which have not yet been clearly delineated or contrasted.
We first want to distinguish quantum control systems which are fully controllable only in the presence of a drift term (i.e., mathematically removing the drift would cause the system to no longer be fully controllable) from those systems for which this is not the case.
Systems of the latter class are known as \emph{strongly controllable} \cite{Elliot}, i.e. they are fully controllable with controls alone regardless of the presence or absence of any drift term. 
In the case of controllable, but not strongly controllable systems, there is an \emph{intrinsic} quantum speed limit.
This is to say, the minimum gate time (over all control fields without any constraints) ultimately has its physical origin in the fact that the implementation of the gate requires exploiting the drift term which is not directly under control, and is bounded in strength.
These speed limits are of the form $T > F(H_0, H_c, U_{g})$. This situation is to be contrasted with systems having a constrained control field $f(t)$, for which the bound on the minimum time arises as a consequence of limitations of the control field, such as bounded amplitude, limited bandwidths, power and energy constraints. 
These bounds are of the form $T>F(H_0, H_c, U_{g},\mathcal F)$ where $\mathcal F$ is the set of admissible controls, and are \emph{extrinsic}, in the sense that they arise not only from limitations on the system Hamiltonian itself, but also from constraints on the control fields.
Typically quantum speed limits in the literature are of the former type since they are not formulated within context of quantum control theory \cite{SpeedLimit1, SpeedLimit2, SpeedLimit4, SpeedLimit5, SpeedLimit6}.
We remark that, contrastingly, every constraint on the overall Hamiltonian $H(t)$ of a controlled quantum system potentially yields a speed limit \cite{SpeedLimit3}. 
However, the latter type of limit remains insufficiently investigated, despite being of critical importance for practical applications of quantum control.
The distinction between intrinsic and extrinsic limits identifies two significantly different types of actionable information in a quantum control scenario.
The extrinsic case indicates when constrained control fields are the limiting factor, whereas the intrinsic limit indicates a physical boundary which cannot be crossed for a given quantum system no matter what type of control is employed. 

The bound derived in this work \eref{eq:lowerboundtime} is of both the intrinsic and extrinsic type.
In the limit $|f_{\text{max}}|\to\infty$, the bound furnishes information purely about the intrinsic speed limit of a given system as the term containing $f_{\text{max}}$ vanishes.
Additionally, in the limit that the term corresponding to the control Hamiltonian goes to zero only the term depending on the drift Hamiltonian persists, and thus the remaining term represents the intrinsic limit.

\section{Conclusions and Outlook}
We have derived a lower bound for the time required to implement a unitary gate through a classical control field.
The bound \eref{eq:lowerboundtime} depends on the strength of the drift and the control Hamiltonian, the highest permitted control field amplitude and the target gate one wants to implement. 
The derived bound can be considered as an extrinsic quantum speed limit since the minimum time to implement a target unitary gate is limited by the maximum control field amplitude and the strength of the internal Hamiltonian.
However, if we allow the control field to be unconstrained, the bound yields an intrinsic quantum speed limit that cannot be crossed, since the speed of the evolution is limited by the norm of the drift Hamiltonian. 

The results in this work are a step towards characterizing the reachable set of gates given a certain evolution time, and thus further establishing a bound on the minimum time $T_{c}$ needed to implement \emph{all} gates. We have provided a criterion for assessing the time $T_c$ at which all gates are reachable in a given system.
This observation has implications for the control landscape \cite{Land1} of the same quantum systems, since the the non-existence of traps (local minima/maxima of the objective functional considered as a function of control(s)) crucially depends on the assumption that the evolution time is sufficiently long to be able to implement all gates (see \cite{Land2} and references therein).
Moreover, using numerical gate optimization, we found that the derived bound is remarkably tight for a single qubit control system; unfortunately this is no longer the case for higher dimensional systems. We argued that this behavior originates from the underlying structure of the dynamical Lie algebra of the control system, particularly the norms of nested commutators and their relation to the desired gate. Furthermore, the interplay between small matrix elements in the drift Hamiltonian and minimum gate times also warrants further investigation.  
 
In this work, and many other works on the quantum speed limit, it is assumed that the time being sought is the minimum time to \emph{perfectly} implement a specific desired unitary operation $U_{g}$, i.e., that the task of interest is to find a pulse which solves $U(T)=U_{g}$.
However, if some error is allowed in the implementation of a gate, does the corresponding minimum gate time defer radically?
The authors conjecture that this is the case for gates within a small neighborhood of any given fast gates, i.e., gates which can be implemented by the controls alone. Furthermore, recently it has been shown that simple analytically obtained pulses can lead to high fidelity gates $\epsilon\approx 0.01$ \cite{Vir1, Vir2}. Further numerical control optimization yields yet higher fidelities at the cost of requiring significantly higher frequency components within the numerically optimized pulse \cite{Vir2}. It would be favorable to obtain criteria characterizing the set of gates for which ``simple'' pulses exist and further to understand the highest frequency required in a pulse to implement a gate perfectly.

\ack
C. A. acknowledges the NSF (grant CHE-1464569) and fruitful discussions with Thomas-Schulte Herbr\"uggen and Robert Zeier. D.B. acknowledges support from the EPSRC Grant No. EP/M01634X/1. B.R. acknowledges the DOE (grant DE-FG-02ER15344) and H. R. acknowledges the ARO (grant W911NF-16-1-0014).

\appendix
\section{Derivation of the inequality \eref{eq:NielsenBound}}
\label{app:derivationNielsen}
Here we verify the inequality \eref{eq:NielsenBound} from the main text based on the work in \cite{Geo1}, which is valid for any unitarily invariant matrix norm, i.e. $\Vert VAU\Vert=\Vert A\Vert $ with $V,U$ being unitary. Consider 
\begin{eqnarray}
\frac{d}{dt}(U_{1}^{\dagger}(t)U_{2}(t))&=U_{1}^{\dagger}(t)(iH_{1}(t))U_{2}(t)+U_{1}^{\dagger}(t)(-iH_{2}(t))U_{2}(t)	\nonumber \\
&=iU_{1}^{\dagger}(t)(H_{1}(t)-H_{2}(t))U_{2}(t),
\end{eqnarray}
such that with $U_{1}(0)=U_{2}(0)=\mathds{1}$ integrating yields
\begin{eqnarray}
U_{1}^{\dagger}(t)U_{2}(t)-\mathds{1} =-i\int_{0}^{t}(U_{1}^{\dagger}(t^{\prime})(H_{1}(t^{\prime})-H_{2}(t^{\prime}))U_{2}(t^{\prime}))dt^{\prime}.	
\end{eqnarray}
We note that for any unitarily invariant norm we have $\Vert U_{2}^{\dagger}(t)U_{1}(t)-\mathds{1} \Vert=\Vert U_{1}(t)-U_{2}(t)\Vert $. Using the triangle inequality and unitary invariance again,
\begin{eqnarray}
\Vert U_{1}(t)-U_{2}(t)\Vert &=\left\Vert -i\int_{0}^{t}(U_{1}^{\dagger}(t^{\prime})(H_{1}(t^{\prime})-H_{2}(t^{\prime}))U_{2}(t^{\prime}))dt^{\prime} \right\Vert  \nonumber\\
\label{eq:usetriangle}
&\leq \int_{0}^{t} \Vert U_{1}^{\dagger}(t^{\prime})(H_{1}(t^{\prime})-H_{2}(t^{\prime}))U_{2}(t^{\prime}) \Vert dt^{\prime} \\ \nonumber 
&=\int_{0}^{t} \Vert H_{1}(t^{\prime})-H_{2}(t^{\prime})\Vert dt^{\prime},
\end{eqnarray}
 we hence arrive at the desired result \eref{eq:NielsenBound}.

\section{Derivation of the lower bound \eref{eq:criticaltimebound}}
\label{app:derivationLB}
In order to derive the lower bound \eref{eq:criticaltimebound} for the time $T_{c}$ at which all gates become reachable we first show that there always exist a $U_{g}\in \text{U}(d)$ with $d$ being even for which 
\begin{eqnarray}
\label{eq:one}
	\Re[\text{tr}\{U_{g}^{\dagger}U^{(i)}\}]\leq \frac{d}{2},~~~~i=1,2,
\end{eqnarray}
holds where $U^{(i)}$ is the unitary evolution generated by the drift and the control Hamiltonian respectively at the time $T_{c}$. Using the eigenbasis $\{\ket{\varphi_{j}}\}_{j=1}^{d}$ of $U_{g}$ with eigenvalues $\exp(-i\lambda_{j})$ respectively the left hand side can be rewritten as 
\begin{eqnarray}
	&\Re[\text{tr}\{U_{g}^{\dagger}U^{(i)}\}]=\sum_{j=1}^{d}\cos(\lambda_{j}+\phi_{j}^{(i)})|\bra{\varphi_{j}}U^{(i)}\ket{\varphi_{j}}| \nonumber \\
	&\leq \frac{d}{2}+\sum_{j~\text{even}~\vee~\text{odd}}\cos(\lambda_{j}+\phi_{j}^{(i)})|\bra{\varphi_{j}}U^{(i)}\ket{\varphi_{j}}|, 
\end{eqnarray}
where $\bra{\varphi_{j}}U^{(i)}\ket{\varphi_{j}}=e^{\phi_{j}^{(i)}}|\bra{\varphi_{j}}U^{(i)}\ket{\varphi_{j}}|$ was used. We observe that $\lambda_{j}$ can always be chosen in such a way that the sum of the right hand side becomes zero. For example take $\lambda_{j}=-\phi_{j}^{(1)}+\frac{\pi}{2}$ for $j$ even and $\lambda_{j}=-\phi_{j}^{(2)}+\frac{\pi}{2}$ for $j$ odd. Thus, for any target evolution $U_{g}$ constructed in this way, inequality \eref{eq:one} holds. Now, for a fully controllable system there always exists a control pulse such that for the time $T_{c}$ the gate $U_{g}$ from above is implemented. Applying  \eref{eq:NielsenBound}, we thus have    
\begin{eqnarray}
&\sqrt{d}\leq \Vert U_{g}-U^{(2)}\Vert \leq T_{c}\Vert H_{0}\Vert \leq T_{c} \sqrt{d}|E_0|,	\\
&\sqrt{d}\leq \Vert U_{g}-U^{(1)}\Vert \leq \int_{0}^{T_{c}}\Vert f(t)H_{c}\Vert\,dt \leq T_{c} \sqrt{d}|f_{\text{max}}E_c|, \nonumber 
\end{eqnarray}
where $E_0$ and $E_c$ are the highest eigenvalues of $H_{0}$ and $H_{c}$, respectively. Combining the bounds from above we hence find \begin{eqnarray}
\frac{1}{2|E_0|}+\frac{1}{2|f_{\text{max}}E_c|}	\leq T_{c}.
\end{eqnarray}

\end{document}